\documentclass[prd,twocolumn,aps]{revtex4}

\usepackage{bm}
\usepackage{amsmath}
\usepackage{amssymb}
\usepackage{latexsym} 
\usepackage{amsfonts} 
\usepackage{epsfig} 

\newcommand{\ket}[1]{\left|#1\right>} 
\newcommand{\bra}[1]{\left<#1\right|} 
 
\newcommand{\nn}{\nonumber\\} 
 
\newcommand{\f}[1]{\mbox{\boldmath$#1$}}

\newcommand{\vau}{\mbox{\boldmath$v$}}
\newcommand{\na}{\mbox{\boldmath$\nabla$}}
\newcommand{\bea}{\begin{eqnarray}}
\newcommand{\ea}{\end{eqnarray}}
\newcommand{\eea}{\end{eqnarray}}
\newcommand{\ord}{{\cal O}}

\begin{document}

\title{On the Universality of the Hawking Effect}
\author{William G.~Unruh$^{1,2}$ and Ralf Sch\"utzhold$^{3}$}
\affiliation{
$^1$Department of Physics \& Astronomy, University of British Columbia, 
6224 Agricultural Road, Vancouver, B.C. V6T 1Z1 Canada;
\\
$^2$Canadian Institute for Advanced Research Cosmology and Gravity Program
\\
$^3$ Institut f\"ur Theoretische Physik, Technische Universit\"at Dresden, 
D-01062 Dresden, Germany
}

\begin{abstract} 
Addressing the question of whether the Hawking effect depends on
degrees of freedom at ultra-high (e.g., Planckian) energies/momenta,
we propose three rather general conditions on these degrees of freedom
under which the Hawking effect is reproduced to lowest order.
As a generalization of Corley's results, we present a rather general
model based on non-linear dispersion relations satisfying these
conditions together with a derivation of the Hawking effect for that
model. 
However, we also demonstrate counter-examples, which do not appear to
be unphysical or artificial, displaying strong deviations from
Hawking's result. 
Therefore, whether real black holes emit Hawking radiation remains an 
open question and could give non-trivial information about Planckian 
physics.  
\end{abstract}
  
\maketitle
 
\section{Introduction}\label{Introduction}

The striking similarity between the laws of black hole physics and the
(zeroth till third) law of thermodynamics motivated the idea to assign 
thermodynamic properties such as temperature and entropy to black
holes \cite{bekenstein}.
Hawking's prediction \cite{hawking} that black holes should emit
thermal radiation with the temperature being consistent with the
thermodynamic interpretation strongly supported this idea.
As a consequence, the concept of black hole entropy as given by the 
surface (horizon) area of the black hole in Planckian units (instead 
of the volume, for example) is now used in many ways to estimate the 
total entropy of other objects -- which is expected to be a measure of 
the number of fundamental degrees of freedom of the underlying theory
(including quantum gravity).

However, in view of the (exponential) gravitational red-shift near the 
horizon, the outgoing particles of the Hawking radiation originate
from modes with extremely large (e.g., trans-Planckian) wavenumbers.
As the known equations of quantum fields in curved space-times are
expected to break down at such wavenumbers, the derivation of the
Hawking radiation has the flaw that it applies a theory beyond its
region of validity. 
This observation poses the question of whether the Hawking effect is  
independent of Planckian physics or not.

One way to address this question is to model the breakdown of the
(usual) local Lorentz invariance (to be expected at the Planck scale)
by a (non-linear) deviation from the linear dispersion relation at
high wavenumbers, see, e.g., \cite{analytical,corley}.
This method is inspired by the black hole analogues which exploit the 
analogy between the propagation of excitations (e.g., sound waves)
in laboratory-physics systems and quantum fields in curved
space-times, see, e.g., \cite{unruh-prl,artificial,droplet}. 

In Sections \ref{Model}, \ref{Analytical}, \ref{Super-luminal}, and 
\ref{Entanglement}, 
we generalize and simplify the model and the results presented by
Corley in \cite{analytical} (see also \cite{corley}) trying to
identify and to present the crucial points.
Section \ref{Universality} is devoted to the question of which
conditions and assumptions regarding Planckian physics are needed to
reproduce Hawking's result -- together with some counter-examples.

\section{Linear Model}\label{Model}

At first we consider a sub-luminal dispersion relation,
cf.~Fig.~\ref{disp-sub}, which is in
some sense conceptually more clear because the in-modes generating the 
Hawking radiation come from outside the black hole.
The horizon acts as a classical turning point where the JWKB
(geometric optics) approximation breaks down allowing phenomena like
particle creation. 
In contrast to Ref.~\cite{analytical}, we shall not specify the shape
of the dispersion relation apart from some rather general assumptions.

\subsection{Wave Equation}

The geometry as seen by the low-energy particles is described in terms of 
the 1+1 dimensional Painlev{\'e}-Gullstrand-Lema{\^\i}tre \cite{pgl} metric
($\hbar=c=1$ throughout)
\bea
\label{pgl}
ds^2
&=&
dt^2-[dx-v(x)\,dt]^2
\nn
&=&
[1-v^2]\,dt^2+2v\,dt\,dx-dx^2
\,.
\ea
The quantity $v(x)$ can be interpreted as the local velocity of the
freely falling frames measured with respect to the time $t$
corresponding to the Killing vector $\partial_t$ of that stationary
metric.  
In terms of the sonic black hole analogues, $t$ is the laboratory time
and $v$ is just the position-dependent velocity of the fluid with the 
(assumed to be constant) speed of sound being absorbed by a
re-definition of the coordinates.
Since the behavior near the horizon in arbitrary dimensions is essentially
1+1 dimensional for each mode, we restrict ourselves to 1+1 dimensions.
Furthermore, we neglect back-scattering (as induced by the angular-momentum
barrier, for example).

\begin{figure}[ht]
\centerline{\mbox{\epsfxsize=8cm\epsffile{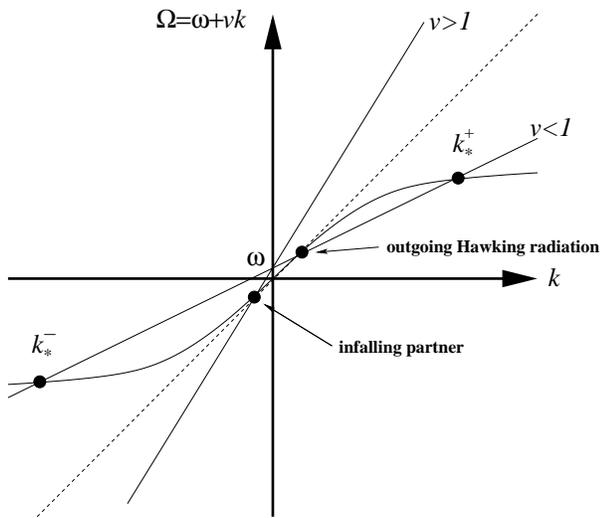}}}
\caption{Sub-luminal dispersion relation (not to scale).
The points of intersection (black circles) with the two lines for
$v>1$ (i.e., $x<0$) and $v<1$ (i.e., $x>0$) determine the solutions of 
the dispersion relation for a given $\omega$. 
The two points corresponding to large wavenumbers $k_*^\pm$ have
group velocities smaller than $v$, i.e., they are "swept away" and
approach the horizon from above $x>0$. 
Hence these solutions are the in-modes. 
The other solution at $x>0$ with the group velocity exceeding $v$
represents the outgoing Hawking radiation. 
The only solution beyond the horizon $x<0$ again has a group velocity
smaller than $v$. 
The corresponding wavefunction represents the infalling partner
particles of the outgoing Hawking radiation, which have a negative
energy as measured from infinity.
During the evolution, the high-wavenumber in-modes $k_*^\pm$ 
($x$ decreases $\leadsto$ $v$ increases) are being converted into the
low-wavenumber Hawking radiation plus partner particles -- 
where the breakdown of the JWKB approximation near the horizon leads 
to a mixing of these modes resulting in particle creation.} 
\label{disp-sub}
\end{figure}

In order to ensure hyperbolicity, causality, and stability, 
we only allow second time-derivatives.
Hence the generalized Klein-Fock-Gordon equation reads
\bea
\label{kfg}
\left(\frac{\partial}{\partial t}+\frac{\partial}{\partial x}\,v(x)\right)
\left(\frac{\partial}{\partial t}+v(x)\,\frac{\partial}{\partial x}\right)
\phi
=
\nn
\left(\frac{\partial^2}{\partial x^2}+
F\left[\frac{\partial^2}{\partial x^2}\right]
\right)\phi
\,,
\ea
with the function $F$ representing the non-trivial dispersion relation.
In general, the function $F$ might contain an arbitrary number of 
derivatives -- i.e., be non-local (think of a lattice, for example).
Note that we do not take into account absorption (i.e., $F$ is real).
The resulting dispersion relation $(\omega+vk)^2=k^2-F[-k^2]$ is
plotted in Fig.~\ref{disp-sub}.

For a stationary metric as in Eq.~(\ref{pgl}), we may separate the most 
general solution of the Klein-Fock-Gordon equation (\ref{kfg}) into 
stationary modes with frequencies $\omega$
\bea
\label{stationary}
&&\left(
F\left[\frac{\partial^2}{\partial x^2}\right]
+[1-v^2]\,\frac{\partial^2}{\partial x^2}+
\right.
\nn
&&\left.
+2v(i\omega-v')\frac{\partial}{\partial x}
-i\omega(i\omega-v')
\right)
\phi_\omega=0
\,.
\ea
The black-hole horizon is assumed to be located at $x=0$ and hence the 
Taylor expansion of the velocity around this point reads
\bea
v(x)=-1+\kappa x+\ord(\kappa^2x^2)
\,,
\ea
with $\kappa$ denoting the surface gravity.

\subsection{Assumptions}

Let us summarize the assumptions that will be used for deriving the Hawking 
effect:

Obviously, the surface gravity of the black hole 
(and hence the temperature of the Hawking radiation)
must be much smaller than the cut-off scale where the concept of geometry 
and metric breaks down (i.e., $F$ is not negligible anymore)
\bea
\kappa \lll k_{\rm cutoff}
\,.
\ea
Furthermore, we shall assume that particle creation -- necessitating a 
break-down of the JWKB (geometric optics) approximation -- occurs in 
the vicinity of the horizon only.
Hence we shall consider an intermediate regime: 
close to the horizon at $x=0$ in units of $\kappa$
\bea
\kappa|x| \ll 1
\,,
\ea
but still many cut-off lengths away from the horizon
\bea
|x|k_{\rm cutoff} \ggg 1
\,.
\ea

Based on the above assumptions, we may neglect terms of second and higher 
order in $\kappa$ and $\omega$ since we are interested in low-frequency 
modes $\omega=\ord(\kappa)$ only (Hawking radiation).
Accordingly, the wave equation (\ref{stationary}) simplifies to 
\bea
\left(
F\left[\frac{\partial^2}{\partial x^2}\right]
+2\kappa\,x\,\frac{\partial^2}{\partial x^2}
-2(i\omega-\kappa)\frac{\partial}{\partial x}
\right)\phi_\omega=0
\,.
\ea
At this stage, it is advantageous to Laplace transform this equation via
\bea
\label{laplace}
\phi_\omega(x)=\int\limits_C ds\,e^{xs}\,\widetilde\phi_\omega(s)
\,.\,
\ea
where the contour $C$ in the complex plane will be discussed below.
Note the change of sign in the second term due to the integration by parts.
The wave equation for the Laplace transformed mode $\widetilde\phi_\omega(s)$
in terms of the complex variable $s$ reads
\bea
\label{lap-kfg}
\left(
F[s^2]-2\kappa\,\frac{\partial}{\partial s}\,s^2
-2(i\omega-\kappa)s\right)\widetilde\phi_\omega=0
\,.
\ea

In the following, we shall impose the following conditions on the 
dispersion relation $F[s^2]$:
\begin{itemize}
\item The dispersion relation $F[s^2]$ is assumed to be an analytic 
function of $s^2$.
\item Hence it possesses a Laurent/Taylor expansion
\bea
\label{laurent}
F[s^2]=k_{\rm cutoff}^2\sum\limits_{n=2}^\infty a_n
\left(\frac{s}{k_{\rm cutoff}}\right)^{2n}
\,,
\ea
where the (non-vanishing) coefficients $a_n$ and the radius of convergence 
are supposed to be of order one -- i.e., the dispersion relation does not 
depend on small quantities like $\kappa/k_{\rm cutoff}$.
\item Furthermore, we assume a sub-luminal dispersion relation, 
i.e., in the rest frame, we have
\bea
\label{sub-luminal}
\left(\frac{d\omega}{dk}\right)^2 \leq 1 
&\leadsto&
0 \leq \omega^2
=
k^2-F[-k^2] \leq k^2
\nn
&\leadsto&
0 \leq F[-k^2] \leq k^2
\,.
\ea
\item Finally, we assume that asymptotically $k^2\uparrow\infty$, 
the dispersion relation is well separated from the line $\omega=k$,
i.e., the phase velocity does not approach unity
\bea
\label{level}
\lim\limits_{k^2\uparrow\infty}\frac{\omega^2}{k^2}
<1
\;\leadsto\;
\lim\limits_{k^2\uparrow\infty}\frac{F[-k^2]}{k^2}
=F_\infty>0
\,.  
\ea
\end{itemize}
Apart from these assumptions we do not need to specify the dispersion
relation any further.
For convenience, we shall choose units in which $k_{\rm cutoff}=1$ 
and omit it in the following equations.

\section{Analytical Derivation}\label{Analytical}

\subsection{Complex Plane and Asymptotics}\label{Complex}

After a separation of variables, the Laplace transformed wave equation 
(\ref{lap-kfg}) can be cast into the following form 
\bea
\frac{\partial}{\partial s}\,\ln\left(s^2\widetilde\phi_\omega\right)=
\frac{F[s^2]-2(i\omega-\kappa)s}{2\kappa\,s^2}
\,.
\ea
Up to an irrelevant pre-factor due to the integration constant, 
its solution reads
\bea
\widetilde\phi_\omega(s)
=
\frac{s^{-i\omega/\kappa}}{s}\,
\exp\left\{
\int ds\,\frac{F[s^2]}{2\kappa\,s^2}
\right\}
\,.
\ea
Since $F[s^2]/s^2$ is analytic, the Laplace transform 
$\widetilde\phi_\omega(s)$ has a singularity at $s=0$ 
and a branch cut from $s=0$ to infinity -- but no 
further singularities at finite values of $s$.
We choose the negative real axis $\Im(s)=0$ and $\Re(s)<0$ for the
branch cut since this choice will be most convenient for deriving
Hawking radiation -- for an alternative choice, see Section
\ref{Entanglement}. 

As in the usual Fourier transform, we choose a contour that approaches 
infinity along the imaginary axis $s=ik$. 
In this case, the overall exponent in Eq.~(\ref{laplace}) is purely 
imaginary and behaves for large $|s|=|k|$ as
\bea
\exp\left\{xs+\int ds\,\frac{F[s^2]}{2\kappa\,s^2}\right\}
\approx
\exp\left\{xs-\frac{sF_\infty}{2\kappa}\right\}
\,,
\ea
according to assumption (\ref{level}).
Hence the exponential function is rapidly oscillating at large $|s|$ and 
thus yields (again at large $|s|$) no contribution to the integral in 
Eq.~(\ref{laplace}).
The $k$-integral over $\exp\{ik[x-F_\infty/(2\kappa)]\}$ gives 
$\delta(2\kappa\,x-F_\infty)$ and hence vanishes since $\kappa|x|\ll1$.
(As we shall see below, the same result can be obtained by deforming the 
contour of integration in the complex plane.)
From a physical point of view, this result is not very surprising since 
-- given a non-trivial dispersion relation -- one would not expect 
momenta which are much larger than the cut-off to contribute.
(This expectation is however false for $F=0$.)
The significant contributions will be found by the stationary phase method, 
or, after deforming the contour, the saddle point method.

\subsection{Saddle-Point Method}\label{Saddle}

In order to apply the saddle point method, let us rewrite the 
Laplace transformation in Eq.~(\ref{laplace}) as
\bea
\phi_\omega(x)=\int\limits_{C}ds\,g(s)\,e^{xf(s)}
\,,
\ea
with the two auxiliary functions 
\bea
g(s)=\frac{s^{-i\omega/\kappa}}{s}
\,,
\ea
and
\bea
f(s)=s+\frac{1}{x}\left(\int ds\,\frac{F[s^2]}{2\kappa\,s^2}\right)
\,.
\ea
The saddle points $s_*$ of $f(s)$ are determined by
\bea
\label{saddle}
\left(\frac{df}{ds}\right)_{s=s_*}=0
\;\leadsto\;
2\kappa\,s^2_*x+F[s^2_*]=0
\,.
\ea
Many cut-off lengths away from the horizon $|x|\ggg1$ 
(but still $\kappa|x|\ll1$), we can approximate the integral in 
Eq.~(\ref{laplace}) by the saddle-point expansion
\bea
\label{expansion}
\phi_\omega(x)\approx\sqrt{\frac{2\pi}{-xf''(s_*)}}\,e^{xf(s_*)}g(s_*)
\,,
\ea
where a sum over multiple saddle points is implied with proper orientation.
The next terms of the saddle-point expansion are suppressed by a factor 
of order 
\bea
\frac{g''(s_*)}{g(s_*)xf''(s_*)}
=
\ord\left(\frac{1}{xs_*}\right)
\,,
\ea
and can be neglected if $x$ is sufficiently large to overcome the smallness 
of $s_*$, which depends on the small quantity $\kappa x$ via 
Eq.~(\ref{saddle}).

Along the imaginary axis $s_*=ik_*$, the possible saddle points are given by 
\bea
F[-k^2_*]=2\kappa\,x\,k^2_*=(1-v^2)k^2_*+\ord(\kappa^2)
\,,
\ea
i.e., they exactly coincide with the solutions of the dispersion
relation (see Fig.~\ref{disp-sub}) for $k\gg\omega$
\bea
k^2_*-F[-k^2_*]=(\omega+vk_*)^2\approx v^2k^2_*
\,,
\ea
since $\omega=\ord(\kappa)$ and $|x| \ggg 1$. 

\subsection{Contour for $x<0$}
%
\begin{figure}[ht]
\centerline{\mbox{\epsfxsize=8cm\epsffile{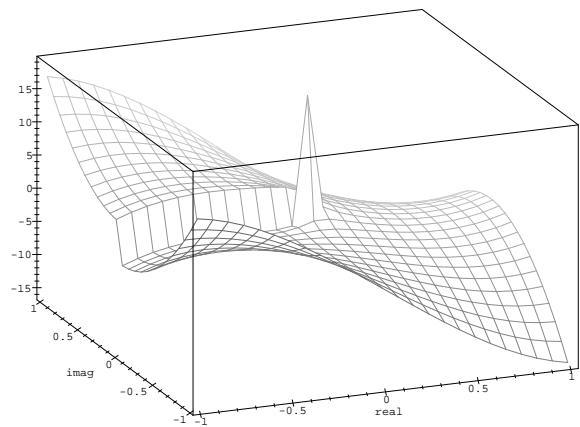}}}
\vspace{.5cm}
\centerline{\mbox{\epsfxsize=7cm\epsffile{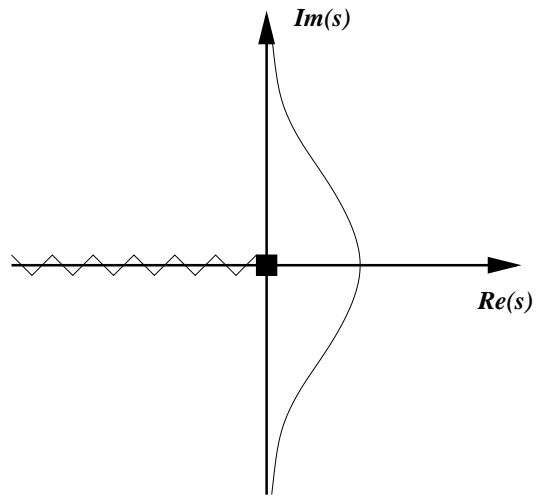}}}
\caption{Landscape plot (top) of the real part of the logarithm of the
integrand in Eq.~(\ref{laplace}) for a sub-luminal dispersion relation
and the case $x=-5$ and $\omega=\kappa=1/30$ as well as contour in the
complex plane (bottom). 
The black square denotes the singularity and the zig-zag line is the
branch cut. 
The behavior of the landscape near the imaginary axis is generic, but
the structure away from that axis 
(e.g., existence of further saddle points)
depends on the particular form of the (sub-luminal) dispersion
relation (here $F[s^2]=s^4$).} 
\label{land1}
\end{figure}

Since the solutions of the dispersion relation and hence the saddle points
depend on the sign of $x$, i.e., on which side of the horizon is considered,
it is convenient to choose different contours in the complex plane for $x>0$
and $x<0$ making sure that they are deformable to each other as $x$ goes 
through zero.
Let us first study the case $x<0$, i.e., the solution of the wave equation 
beyond the horizon, cf.~Fig.~\ref{land1}.

Along the imaginary axis $s=ik$ the exponent in Eq.~(\ref{expansion}) is 
purely imaginary 
\bea
\Re\{f(ik)\}=0
\,,
\ea
whereas its derivative is purely real and positive
\bea
f'(ik)=1-\frac{F[-k^2]}{2\kappa\,x\,k^2}>0
\,,
\ea
because $x<0$ and $F[-k^2]\geq0$, cf.~assumption (\ref{sub-luminal}).
As a result $|\exp\{xf(s)\}|$ decreases rapidly ($x<0$) with increasing 
$\Re\{s\}$ and there are no saddle points on the imaginary axis $s=ik$.
Hence we can deform the contour into the valley at $\Re\{s\}>0$
until we hit possible saddle points at 
\bea
\Re\{s_*[x<0]\}>0
\,,
\ea
with values 
\bea
\Re\{f(s_*[x<0])\}>0
\,.
\ea
Since the coefficients in the Laurent/Taylor expansion of $F[s^2]$ are 
of order one, the real part of the saddle point satisfying 
$2\kappa\,s^2_*x+F[s^2_*]=0$ is mainly determined by the small quantity 
$\kappa\,x\ll1$.
Again we assume that the size of $x\ggg1$ overcomes the smallness of 
$\kappa\,x\ll1$ and consequently obtain a solution 
\bea
\phi_\omega(x<0)\approx\sqrt{\frac{2\pi}{-xf''(s_*)}}\,e^{xf(s_*)}g(s_*)
\,,
\ea
which decays exponentially fast beyond the horizon.

Note that, even if we encounter no saddle points, the contour can still 
be deformed such that the contributions are exponentially small, 
cf.~Fig.~\ref{land1}.
In this way the chosen contour yields basically no contribution beyond the 
horizon $x<0$ -- which is exactly what we want for the derivation of the 
outgoing Hawking radiation.

\subsection{Contour for $x>0$}
%
\begin{figure}[ht]
\centerline{\mbox{\epsfxsize=8cm\epsffile{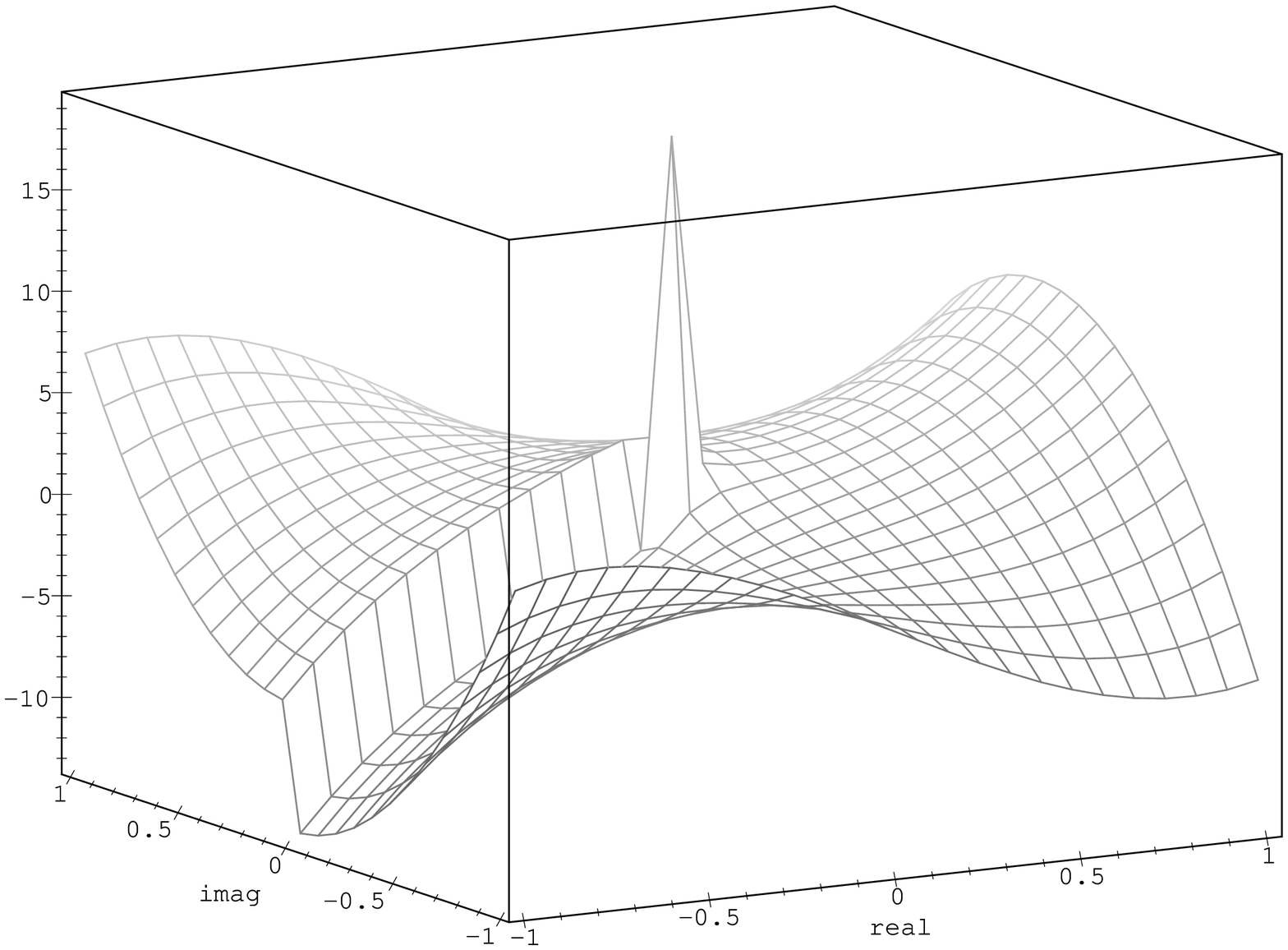}}}
\vspace{.5cm}
\centerline{\mbox{\epsfxsize=7cm\epsffile{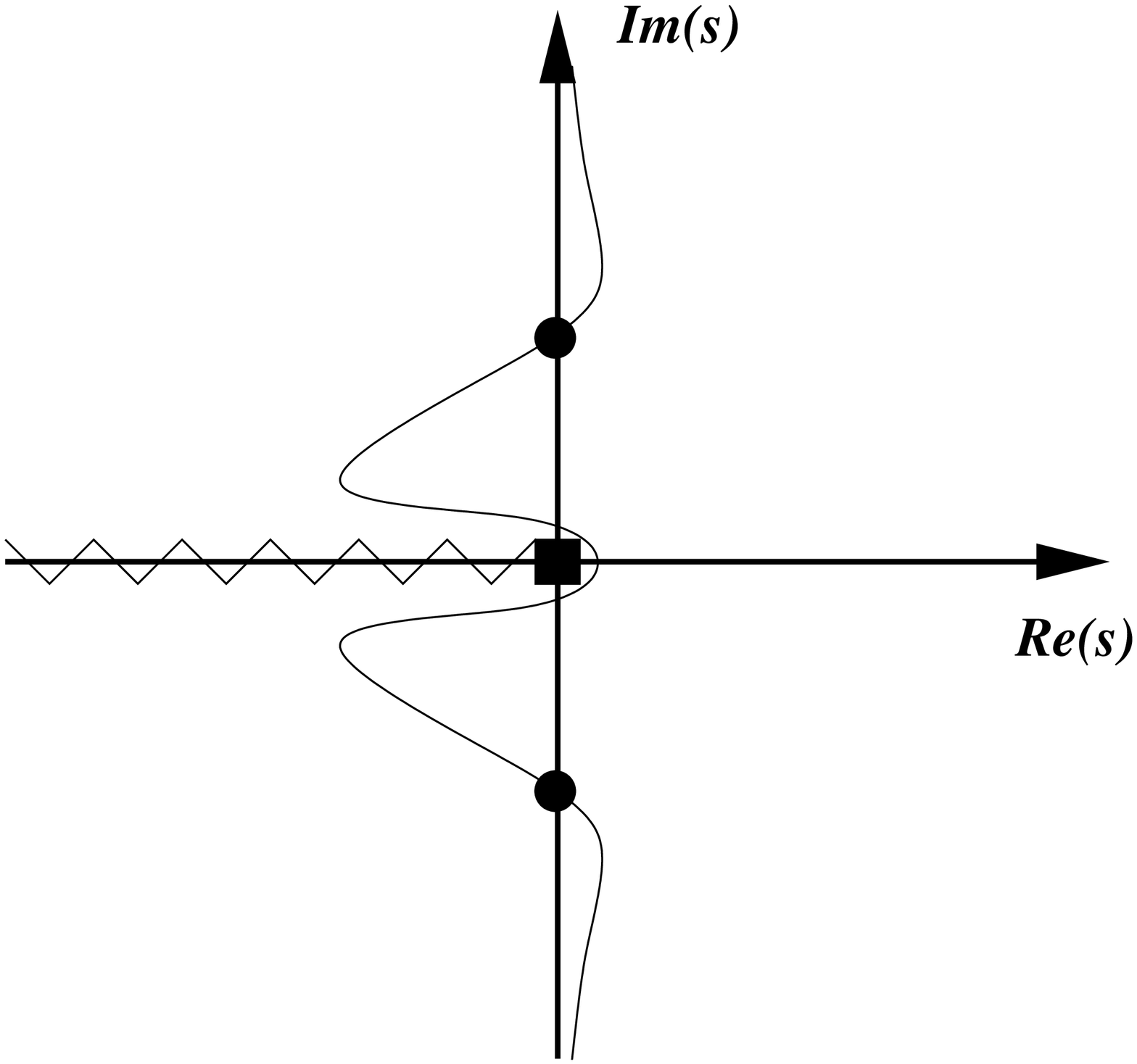}}}
\caption{Landscape plot (top) of the real part of the logarithm of the
integrand in Eq.~(\ref{laplace}) for a sub-luminal dispersion relation
and the case $x=+5$ and $\omega=\kappa=1/30$ as well as contour in the
complex plane (bottom). 
The black square denotes the singularity, the black dots are the
saddle points, and the zig-zag line is the branch cut.
The behavior of the landscape near the imaginary axis is generic, but
the structure away from that axis 
(e.g., existence of further saddle points) 
depends on the particular form of the (sub-luminal) dispersion
relation (here $F[s^2]=s^4$).}
\label{land2}
\end{figure}

In order to derive the solution outside the horizon $x>0$, another contour
is needed for applying the saddle-point method, cf.~Fig.~\ref{land2}.
The exponent in Eq.~(\ref{expansion}) is still purely imaginary along the 
imaginary axis $s=ik$, but the slope  
\bea
f'(ik)=
1-\frac{F[-k^2]}{2\kappa\,x\,k^2}
\,,
\ea
changes its sign at saddle points $s_*^\pm=\pm ik_*$.
The condition (\ref{laurent}) ensures the existence of exactly two 
(symmetric) saddle points along the imaginary axis -- i.e., solutions 
of the dispersion relation with finite values of $f''(s_*)$ -- but the 
analysis can easily be generalized to the case of more than two saddle 
points.

For small $k\ll1$, the first term dominates according to Eq.~(\ref{laurent})
\bea
F[-k^2\uparrow0] \ll k^2
\;\leadsto\;
1-\frac{F[-k^2]}{2\kappa\,x\,k^2} \approx 1
\,,
\ea
and hence the valley is on the side $\Re\{s\}<0$ of the imaginary axis -- 
whereas for $k^2>k_*^2$, the slope $f'(ik)$ is negative and thus the valley 
is on the other side $\Re\{s\}>0$.
Hence the contour must cross these two saddle points (cf.~Fig.~\ref{land2})
and pick up the corresponding contributions
\bea
\label{phi-pm}
\phi_\omega^\pm(x>0)
&\approx&
\sqrt{\frac{2\pi}{-xf''(s_*^\pm)}}\,e^{xf(s_*^\pm)}g(s_*^\pm)
\nn
&=&
\sqrt{\frac{2\pi}{-xf''(\pm ik_*)}}\,e^{xf(\pm ik_*)}g(\pm ik_*)
\nn 
&=&
\sqrt{\frac{2\pi}{\mp xf''(ik_*)}}\,e^{\pm xf(ik_*)}g(\pm ik_*)
\,.
\ea
As $f(ik_*)$ is purely imaginary, the only difference in the absolute
values of the two contributions is determined by the branch cut in $g(s)$
\bea
\label{ratio-cut}
\left|\frac{\phi_\omega(s_*^+,x>0)}{\phi_\omega(s_*^-,x>0)}\right|=
\left|\frac{g(s_*^+)}{g(s_*^-)}\right|=e^{\pi\omega/\kappa}
\,.
\ea
Ergo, the two saddle points at $s_*^\pm=\pm ik_*$ yield two rapidly 
($x\ggg1$) and oppositely oscillating contributions, whose absolute
values satisfy the above relation (which will become important later on).

\subsection{Branch Cut}\label{Branch}

Between the two saddle points $-k_*<k<k_*$, the valley lies on the 
same side $\Re\{s\}<0$ of the imaginary axis as the branch cut does,
cf.~Fig.~\ref{land2}.
If it was not for the branch cut, the contour (originating from infinity)
could be closed in this valley after crossing the two saddle points at 
$s_*^\pm=\pm ik_*$ such that all additional contributions 
(possibly further saddle points) are exponentially smaller than those 
of the saddle points at $s_*^\pm=\pm ik_*$.
However, the branch cut demands that we integrate along it to $s=0$ 
from both sides (with the proper orientation) and in circumventing 
the branch cut, we pick up the difference in the values of $g(s)$ 
\bea
g(\Im(s)\downarrow0)-g(\Im(s)\uparrow0)=
2\sinh\left(\frac{\pi\omega}{\kappa}\right)
\frac{|s|^{-i\omega/\kappa}}{s}
\,.
\ea
In this way, we obtain an additional contribution
\bea
\phi_\omega^{\rm rest}(x)
&=&
2\sinh\left(\frac{\pi\omega}{\kappa}\right)
\int\frac{ds}{s}\,
|s|^{-i\omega/\kappa}
\times
\nn
&&\times
\exp\left\{xs+\int ds\,\frac{F[s^2]}{2\kappa\,s^2}\right\}
\,,
\ea
where the integral runs from $0$ along the negative real axis up to
the intersection point of the contour with the branch cut.
In view of $x\ggg1$, we can omit the second term in the integrand and
extend the interval to $-\infty$
\bea
\phi_\omega^{\rm rest}(x>0)\approx
2\sinh\left(\frac{\pi\omega}{\kappa}\right)
\int\limits_0^{-\infty}\frac{ds}{s}\,
\frac{\exp\left\{xs\right\}}{|s|^{i\omega/\kappa}}
\,.
\ea
Outside the horizon $x>0$ -- the region we are interested in -- 
we may substitute $\chi=|xs|$ and obtain
\bea
\phi_\omega^{\rm rest}(x>0)\approx
-2\sinh\left(\frac{\pi\omega}{\kappa}\right)
x^{i\omega/\kappa}
\int\limits_0^{\infty}\frac{d\chi}{\chi}\,
\frac{e^{-\chi}}{\chi^{i\omega/\kappa}}
\,,
\ea
which is just an integral representation of the $\Gamma$-function, i.e.,
\bea
\phi_\omega^{\rm rest}(x>0)\approx
-2\sinh\left(\frac{\pi\omega}{\kappa}\right)
\Gamma\left(-\frac{i\omega}{\kappa}\right)
\,x^{i\omega/\kappa}
\,.
\ea
Together with the contributions in Eq.~(\ref{phi-pm}), this completes the 
(approximate) solution of the wave equation outside the horizon -- 
for the case that the solution basically vanishes beyond the horizon.
Of course, we can only draw this conclusion if the two contours for $x>0$
and $x<0$ in Figs.~\ref{land1} and \ref{land2} are deformable to each
other as $x$ crosses zero. 
This property is ensured by assumption (\ref{laurent}) since the part of the 
complex plane covered during the deformation of the contours 
($\kappa|x|\ll1\,\leadsto\,|s_*|\ll1$) is well inside the radius of 
convergence of order one.

\subsection{Bogoliubov Coefficients}\label{Bogoliubov}

Let us identify the various parts of the solution.
The contribution generated by the branch cut $\phi_\omega^{\rm rest}(x)$
is the wavefunction of an outgoing particle with a low wavenumber
$k=\ord(\omega)$ (e.g., Hawking radiation).
The saddle-point contributions $\phi_\omega^\pm(x)$, on the other hand,
are rapidly oscillating, 
since the largeness of $x\ggg1$ is supposed to be stronger than the
smallness of $s_*(\kappa x)$. 

As one can observe in Fig.~\ref{disp-sub}, the group velocity of the
low-energy 
mode $\phi_\omega^{\rm rest}(x)$ exceeds $v(x)$, as one should expect for 
an outgoing particle -- whereas the group velocity of the rapidly 
oscillating modes $\phi_\omega^\pm(x)$ is smaller than $v(x)$.
Hence these are the in-modes $\phi_\pm^{\rm in}$.

Furthermore, the frequencies of the rapidly oscillating modes in the freely 
falling frame $\Omega_\pm=\omega\pm vk_*$ have different signs because
$k_*\gg\omega$ (although $\omega>0$ for both modes).
As a result, the low-energy outgoing particle (e.g., Hawking radiation)
is a mixture of positive and negative frequency 
(with respect to the freely falling frame) in-modes -- 
which can be described in terms of the Bogoliubov coefficients
\bea
\label{bogol}
\phi_\omega^{\rm out}=\alpha_\omega\phi_+^{\rm in}+\beta_\omega\phi_-^{\rm in}
\,.
\ea
A non-vanishing Bogoliubov $\beta_\omega$-coefficient of course corresponds to
the phenomenon of particle creation -- i.e., the in-vacuum with
respect to the freely 
falling frame is converted into a quantum state containing particles 
(Hawking radiation) by the horizon.
The ratio of the Bogoliubov coefficients is determined by 
Eq.~(\ref{ratio-cut})
\bea
\label{balance}
|\beta_\omega|=e^{-\pi\omega/\kappa}|\alpha_\omega|
\,,
\ea
which is the well-known relation leading to the thermal Hawking spectrum.
E.g., applying the unitarity relation of the Bogoliubov coefficients
$\f{\alpha}\cdot\f{\alpha}^\dagger-\f{\beta}\cdot\f{\beta}^\dagger=\f{1}$,
we immediately obtain the thermal spectrum 
$\langle\hat N_\omega\rangle=|\beta_\omega|^2\propto
1/(e^{2\pi\omega/\kappa}-1)$.

Note that the quantum state generated by the time-evolution of the
in-vacuum  also contains particles with low wavenumbers beyond the
horizon (see Fig.~\ref{disp-sub}). 
As required by energy conservation and unitarity, for each outgoing particle 
of the Hawking radiation, there is a corresponding partner particle with 
negative energy (as measured from infinity) inside of the black hole -- 
but this does not alter presented calculation, cf.~\cite{analytical}.
There are, however, correlations between the outgoing Hawking particle 
and its partner beyond the horizon -- which generate the true thermal 
character (mixed state instead of pure state) of the Hawking radiation
for any outside observer 
(thermo-field formalism \cite{corley,thermo}), 
see Section \ref{Entanglement} below.

The branch cut in the complex plane caused by the horizon turns out to be
a main ingredient for deriving the Hawking effect -- it generates the 
contribution $\phi_\omega^{\rm rest}(x)$ as well as the ratio in 
Eq.~(\ref{ratio-cut}) -- which are both essential features.

\section{Super-luminal Dispersion}\label{Super-luminal}
%
\begin{figure}[ht]
\centerline{\mbox{\epsfxsize=8cm\epsffile{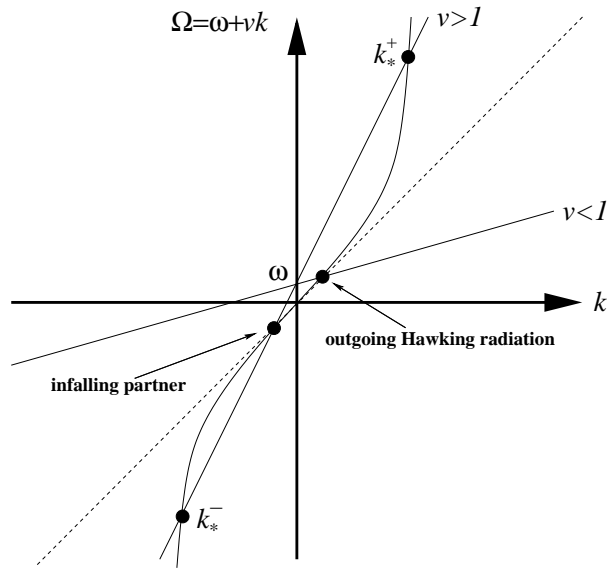}}}
\caption{Super-luminal dispersion relation (not to scale). 
The points of intersection (black circles) with the two lines for
$v>1$ (i.e., $x<0$) and $v<1$ (i.e., $x>0$) determine the solutions of
the dispersion relation for a given $\omega$.
As in the sub-luminal case, the out-modes are the low-wavenumber
solutions corresponding to the outgoing particles of the Hawking
radiation ($x>0$) and their infalling partners ($x<0$).
However, the high-wavenumber in-modes $k_*^\pm$ have group velocities
exceeding $v$ and hence are approaching the horizon from the inside
$x<0$.}
\label{disp-sup}
\end{figure}

So far, we restricted our attention to a sub-luminal dispersion
relation only. 
As we shall see now, the case of a super-luminal dispersion as in
Fig.~\ref{disp-sup} can be treated in basically the same way.
The steps and derivations from Eq.~(\ref{pgl}) to Eq.~(\ref{laurent}) 
are identical, and we choose the same branch cut.
Of course, for a super-luminal dispersion relation we have to modify 
assumptions (\ref{sub-luminal}) and (\ref{level}) accordingly in order
to ensure a vanishing asymptotical contribution at the imaginary axis $s=ik$.
The derivations in Sections \ref{Complex} and \ref{Saddle} apply in the same 
way, but now the solutions of the dispersion relation with large wavenumbers
(the in-modes) are super-luminal, 
i.e., they originate from inside the black hole, see Fig.~\ref{disp-sup}.

\subsection{Contour for $x>0$}
%
\begin{figure}[ht]
\centerline{\mbox{\epsfxsize=8cm\epsffile{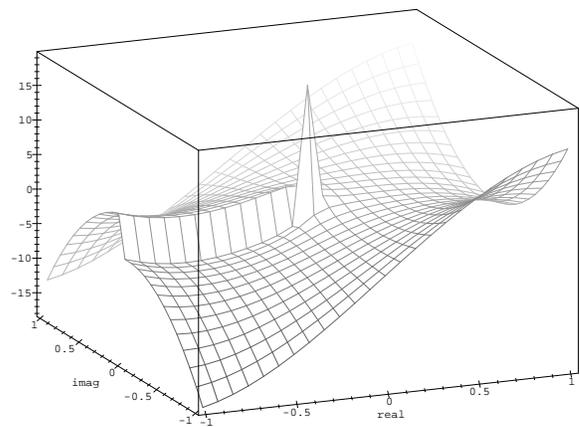}}}
\vspace{.5cm}
\centerline{\mbox{\epsfxsize=8cm\epsffile{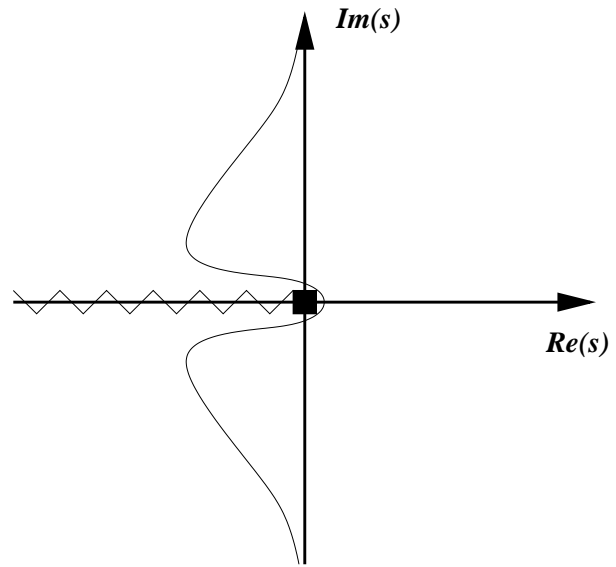}}}
\caption{Landscape plot (top) of the real part of the logarithm of the
integrand in Eq.~(\ref{laplace}) for a super-luminal dispersion
relation and the case $x=+5$ and $\omega=\kappa=1/30$ as well as
contour in the complex plane (bottom). 
The black square denotes the singularity and the zig-zag line is the
branch cut. 
The behavior of the landscape near the imaginary axis is generic, but
the structure away from that axis 
(e.g., existence of further saddle points) 
depends on the particular form of the (super-luminal) dispersion
relation (here $F[s^2]=-s^4$).}
\label{land3}
\end{figure}

Let us first consider the solution outside the black hole, see
Fig.~\ref{land3}. 
In contrast to the sub-luminal case, the function $f(s)$ in the exponent 
has a positive slope for all $k$-values
\bea
f'(ik)=1-\frac{F[-k^2]}{2\kappa\,x\,k^2}>0
\,,
\ea
because $F[-k^2]<0$, and, consequently, the valley is now situated at 
negative real parts of $s$.
Deforming the contour into the valley, all contributions become exponentially 
small -- but (again) we have to circumvent the branch cut.
The contribution of the branch cut yields the same result as in 
Section \ref{Branch}.
Ergo, outside the black hole, we have only the outgoing Hawking particle --
which is exactly what one would expect in the super-luminal case.

\subsection{Contour for $x<0$}
%
\begin{figure}[ht]
\centerline{\mbox{\epsfxsize=8cm\epsffile{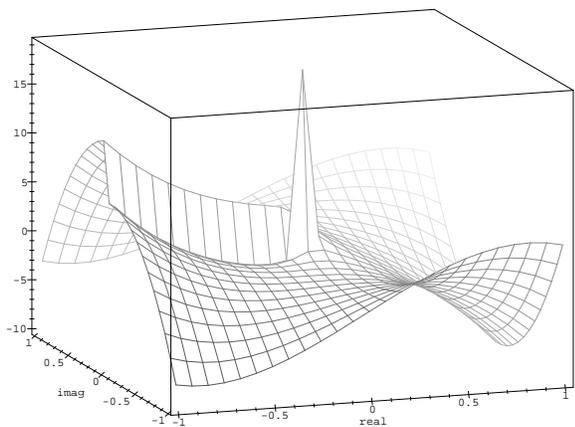}}}
\vspace{.5cm}
\centerline{\mbox{\epsfxsize=8cm\epsffile{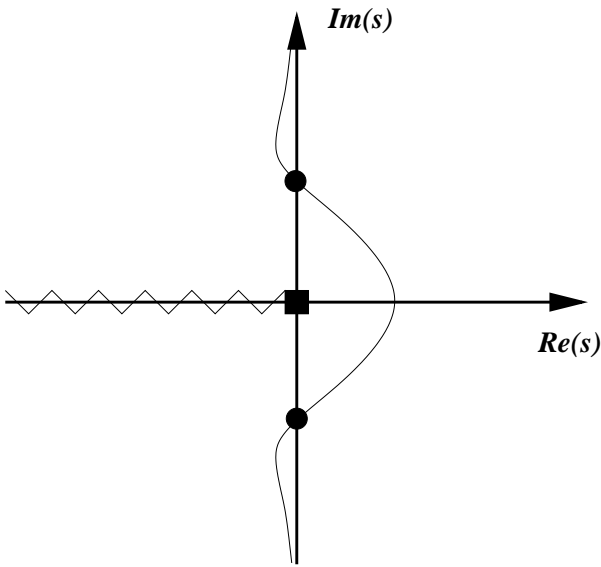}}}
\caption{Landscape plot (top) of the real part of the logarithm of the
integrand in Eq.~(\ref{laplace}) for a super-luminal dispersion
relation and the case $x=-5$ and $\omega=\kappa=1/30$ as well as
contour in the complex plane (bottom).
The black square denotes the singularity, the black dots are the
saddle points, and the zig-zag line is the branch cut.
The behavior of the landscape near the imaginary axis is generic, but
the structure away from that axis 
(e.g., existence of further saddle points) 
depends on the particular form of the (super-luminal) dispersion
relation (here $F[s^2]=-s^4$).}
\label{land4}
\end{figure}

For $x<0$, i.e., beyond the horizon, the slope $f'(ik)$ changes its sign at 
the saddle points 
(i.e., the solutions of the dispersion relation with large wavenumbers)
and the contour has to cross the imaginary axis picking up the
saddle-point contributions, cf.~Fig.~\ref{land4}. 
For those contributions, we basically obtain the same results as in 
Eq.~(\ref{phi-pm}) and hence in Eq.~(\ref{ratio-cut}), because the 
branch cut is identical.

Therefore, we reach the same conclusion as in Section \ref{Bogoliubov} 
where now the in-modes originate from inside the black hole.
Hence we reproduce Hawking radiation also for a super-luminal dispersion 
-- provided that the in-modes (large wavenumbers) are initially in their 
ground state with respect to the freely falling frame.


\section{Entanglement}\label{Entanglement}

So far, we restricted our attention to the decomposition of the
outgoing Hawking radiation in terms of the in-modes,
cf.~Eq.~(\ref{bogol}).  
However, a full description of the evolution of the quantum state
requires a complete set of out-modes, i.e., the outgoing Hawking
particles as well as their infalling partners. 
Fortunately, it turns out that the Bogoliubov coefficients of the
infalling partners can be inferred in complete analogy to the previous
Sections if we choose the branch cut in the opposite way, i.e., along
the positive real axis $\Im(s)=0$ and $\Re(s)>0$.
Circumventing this branch cut then reproduces the wavefunction of the
infalling partner particles as in Sec.~\ref{Branch} 
(but with $x\to-x$), and, consistently, this contribution only
occurs for $x<0$, i.e., beyond the horizon.
Again, for both cases (sub- and super-luminal), one obtains basically the
same relation 
\bea
\phi_\omega^{\rm partner}=
\beta_\omega^{\rm partner}\phi_+^{\rm in}
+
\alpha_\omega^{\rm partner}\phi_-^{\rm in}
\,.
\ea
Note that we have interchanged the role of the creation and
annihilation operators and hence $\alpha^{\rm partner}$ and
$\beta^{\rm partner}$ here because the energy of the infalling partner 
particles is negative as measured from infinity and their pseudo-norm
is negative for positive $\omega$.
The opposite direction of the branch cut implies the inverse relation
compared to Eq.~(\ref{ratio-cut}) and together with the above
interchange, we derive the same ratio as in Eq.~(\ref{balance})
\bea
\label{partner}
|\beta_\omega^{\rm partner}|=
e^{-\pi\omega/\kappa}|\alpha_\omega^{\rm partner}|
\,.
\ea
The knowledge of the complete set of out-modes 
(Hawking radiation $\phi_\omega^{\rm Hawking}$ plus their infalling
partners $\phi_\omega^{\rm partner}$)
facilitates the decomposition of the in-modes in terms of the out-modes    
\bea
\phi_-^{\rm in}
=
\alpha_\omega^{\rm inv}\phi_\omega^{\rm partner}+
\beta_\omega^{\rm inv}\phi_\omega^{\rm Hawking}
\,.
\ea
In view of the relations (\ref{balance}) and (\ref{partner}) as well
as unitarity
$\f{\alpha}\cdot\f{\alpha}^\dagger-\f{\beta}\cdot\f{\beta}^\dagger=\f{1}$,
the inverse Bogoliubov coefficients satisfy an analogous condition 
\bea
|\beta_\omega^{\rm inv}|=e^{-\pi\omega/\kappa}|\alpha_\omega^{\rm inv}|
\,.
\ea
Consequently, the in-vacuum defined via 
$\hat a_{\rm in}\ket{0_{\rm in}}=0$
will be annihilated by a linear combination of the operators
corresponding to the out-modes 
\bea
\left[\hat a_\omega^{\rm partner}+e^{-\pi\omega/\kappa}
(\hat a_\omega^{\rm Hawking})^\dagger\right]\ket{0_{\rm in}}=0
\,,
\ea
where an irrelevant phase has been absorbed by the re-definition of
$\hat a_\omega^{\rm partner}$.  
This well-known relation (see, e.g., \cite{unruh-prd})
induces the entanglement between the particles of the Hawking
radiation $\hat a_\omega^{\rm Hawking}$ and their infalling partners
$\hat a_\omega^{\rm partner}$
\bea
\ket{0}_{\rm in}^\omega
\propto
\exp\left\{
e^{-\pi\omega/\kappa}
\left(
\hat a_\omega^{\rm partner}\,\hat a_\omega^{\rm Hawking}
\right)^\dagger
\right\}\ket{0}_{\rm out}^\omega
\,,
\ea
which in turn generates the thermal density matrix after averaging
over the unobservable infalling partners.


\section{Universality}\label{Universality}

The derivation presented in the previous Sections demonstrates that the 
the Hawking effect does (to lowest order) not depend on the details of 
the dispersion relation at high wavenumbers -- given the model 
assumptions discussed above. 
Let us try to identify more general conditions under which the Hawking should
remain unchanged by the details of the physics at large wavenumbers.
For convenience, we shall assume that the cut-off scale coincides with the 
Planck scale and use the terms sub-Planckian for effects according to the 
known laws of physics (e.g., linear dispersion) and trans-Planckian for 
new physics (e.g., non-linear dispersion).

First of all, we assume that the JWKB (geometric optics or eikonal) 
approximation breaks down 
(thereby allowing for the phenomenon of particle creation) 
in the vicinity of the horizon only, where the gravitational red-shift
induces a transition of trans-Planckian into sub-Planckian modes.
An example where this assumption does not apply will be discussed in 
Section \ref{Breakdown}.

Given that assumption, the crucial point is the quantum state of the 
modes when they leave the Planckian regime.
If the modes leave the Planckian regime (''are born'') in their ground
state with respect to freely falling observers near the horizon, then 
one obtains Hawking radiation, cf.~\cite{unruh-prd,haag}.
Let us review the standard argument leading to that conclusion.
In terms of the Regge-Wheeler tortoise coordinate $r_*$, the 1+1 dimensional 
Schwarzschild metric can be cast into the conformally flat form 
\bea
\label{1+1-schwarz}
ds^2
&=&
\left(1-\frac{2M}{r}\right)\left(dt^2-dr_*^2\right)
\nn
&\simeq&
\exp\left\{\frac{r_*}{2M}\right\}(dt^2-dr^2_*)
\,,
\ea
where the $\simeq$ applies near the horizon.
The trajectory of a freely falling observer is given by 
($A$ and $B$ are integration constants)
\bea
\label{trajectory}
r_*(t\uparrow\infty)\simeq-t-A\,\exp\left\{-\frac{t}{2M}\right\}+B
\,,
\ea
and its proper time $d\tau^2=ds^2[t,r_*(t)]$ accordingly reads
\bea
\label{proper}
\tau\sim\exp\left\{-\frac{t}{2M}\right\}
\,.
\ea
Hence the freely falling observers would define their ground state 
via the positive frequency solutions 
\bea
F^{\rm in}_\omega(U)=\frac{1}{\sqrt{4\pi\omega}}\,
e^{-i\omega U}
\,,
\ea
with respect to the Kruskal coordinate 
\bea
U=-4M\,e^{-u/(4M)}=-4M\,e^{-[t-r_*]/(4M)}
\,.
\ea
The doubly exponential behavior of these modes -- 
when expressed in terms of the coordinates $t,r$ of an outside observer 
-- lead to the thermal particle content.

The remaining issue is, of course, to determine in which cases the modes 
do indeed leave the Planckian regime in their ground state with respect 
to freely falling observers (near the horizon).
As a very natural example, one could ensure this property by means of the 
following three assumptions:
\begin{itemize}
\item[{\bf a)}] {\bf Freely falling frame}\\
If we assume that the usual local Lorentz invariance is broken at the 
Planck scale via the introduction of preferred frames 
(where preferred frames are the frames in which Planckian physics 
displays maximal symmetry under time-inversion, for example)
then the freely falling frame should be preferred 
(instead of the rest frame of the black hole, for example).
\item[{\bf b)}] {\bf Ground state}\\
The Planckian excitations are assumed to start off in their ground state
(with respect to the freely falling frame, see point above) subject to 
possible constraints such as conservation laws etc.
\item[{\bf c)}] {\bf Adiabatic evolution}\\
Finally the evolution of the modes is supposed to be adiabatic -- i.e.,
the Planckian dynamics is supposed to be much faster than all external 
(sub-Planckian) variations (e.g., experienced by a traveling wavepacket).
This condition demands the absence of level crossing and long time-scales 
in Planckian physics.
\end{itemize}

E.g., for the sonic black hole analogues (''dumb holes'') such as  
a fluid flowing trough a Laval nozzle (accelerated from sub-sonic 
to super-sonic speed),
the freely falling frame corresponds to the local rest frame of the 
flowing fluid, whereas the rest frame of the walls of the nozzle 
is analogous to the global rest frame of black hole.
Of course, one can easily imagine situations where at least one of the 
above assumptions fails.
E.g., for a super-luminal dispersion relation, the modes with large 
wavenumbers originate from inside the black hole, i.e., ultimately
from the singularity (of from a turbulent regime), and it is not 
obvious why they should be in their ground state.
Further examples for the failure of the above assumptions, where the 
reference frame for Planckian physics is not the local freely falling 
frame but the global rest frame of black hole; or where the adabaticity
breaks down, are the subject of the next Sections.

\subsection{Miles Instability}\label{Miles}

As an example, in which the aforementioned set of assumptions fails and 
which does not reproduce Hawking radiation, let us consider the following
fluid model:
Apart from a deviation from the usual dispersion relation as described by
a $k$-dependent phase velocity $v_{\rm ph}^2(k)$, we suppose a coupling to 
a reservoir of Planckian degrees of freedom in the rest frame of the black 
hole (i.e., {\em not} the freely falling frame) manifesting itself as an
effective dissipation term in the dispersion relation
\bea
\label{damping}
(\omega+v_{\rm fl}k)^2=
k^2v_{\rm ph}^2(k)-2i\omega\gamma(k)
\,.
\ea
The $k$-dependence of the damping term $\gamma(k)$ ensures that it is 
completely negligible at sub-Planckian wavenumbers.

In flat space-time (fluid at rest $v_{\rm fl}=0$), the damping term just
implies a decay of the Planckian modes (as one would expect).
For a black hole, the Planckian modes giving rise to Hawking radiation, 
however, behave in a different way.
As one can easily perceive from Figs.~\ref{disp-sub} and
\ref{disp-sup}, for solutions of the dispersion  
relation with large (Planckian) wavenumbers, we have
\bea
\left|v_{\rm ph}^2(k)-v_{\rm fl}^2\right|\ll1
\,.
\ea
Expanding the relevant solution of Eq.~(\ref{damping}) for $\omega$ 
in powers of this small quantity, we obtain
\bea
\omega_+\approx\frac{k^2}{2}\,
\frac{v_{\rm ph}^2(k)-v_{\rm fl}^2}{v_{\rm fl}k+i\gamma(k)}
\,.
\ea
Assuming a real wavenumber, the imaginary part of the frequency changes its 
sign if the fluid velocity exceeds the phase velocity 
\bea
\Im(\omega)>0
\,,
\ea
which indicates an instability.
This phenomenon is basically the Miles instability -- which is responsible 
for the generation of water waves by wind, for example \cite{miles}.

If $k$ and $\gamma$ are of order one (in Planckian units), the imaginary part 
of $\omega$ is of the same order as the real part
\bea
\Im(\omega)=\ord[\Re(\omega)]
\,,
\ea
and since $\omega=\ord(\kappa)$ corresponds to the inverse size of the 
black hole, there can be enough time for the instability to develop
and to excite the modes.
Note that positive frequency (trans-Planckian) modes with 
$v_{\rm ph}^2(k)>v_{\rm fl}^2$ are damped but negative frequency modes 
with $v_{\rm ph}^2(k)<v_{\rm fl}^2$ are amplified.
Hence this effect destroys the balance in Eq.~(\ref{balance}) which 
generates the thermal spectrum of the Hawking radiation.

However, the above analysis based on classical solutions of the dispersion
relation cannot be applied directly, i.e., without respecting the 
fluctuation-dissipation theorem, for example, to the quantum fluctuations 
that generate the Hawking radiation.
In order to turn our attention to the quantum theory, let us consider the 
following Lagrangian density corresponding to a super-luminal dispersion
$v_{\rm ph}^2=1+k^2$
\bea
{\cal L}
=
\frac{1}{2}\,\left[
(\dot\phi+\vau\cdot\na\phi)^2
-(\na\phi)^2
-(\na^2\phi)^2
\right]
\,.
\ea
For a stationary metric, i.e., $v(x)$, a conserved energy density with 
respect to global rest frame of black hole can be derived by means of 
the Noether theorem 
\bea
{\cal E}
=
\frac{1}{2}\,\left[
\dot\phi^2
+(\na\phi)^2
+(\na^2\phi)^2
-(\vau\cdot\na\phi)^2
\right]
\,.
\ea
Evidently the energy density is not positive definite for 
$\vau^2>v_{\rm ph}^2$, i.e., beyond the horizon (super-luminal dispersion).
The local energy density with respect to the local freely falling frame is 
of course positive definite. 
After a normal mode expansion into wavepackets, the total Hamiltonian is 
split up into nearly independent positive and negative energy modes 
(with respect to global rest frame of black hole). 
Obviously, the negative energy modes can be strongly excited by a comparably
weak interaction with further Planckian degrees of freedom at the global rest 
frame of the black hole.
In this way, these modes would not be in their ground state -- even with 
respect to freely falling observers -- and, consequently, one would not 
reproduce Hawking radiation.

For example, let us consider fermionic fields where the quantum states of 
all trans-Planckian modes are maximally excited, i.e., $\ket{1}$ instead 
of the ground state $\ket{0}$.
In that situation, the usual relation 
$\beta_\omega=e^{-\pi\omega/\kappa}\alpha_\omega$ 
implies
\bea
\bra{1}\hat N_\omega\ket{1}
=
|\alpha_\omega|^2
\propto
\frac{1}{1+e^{-2\pi\omega/\kappa}}
\,,
\ea
i.e., {\em not} a thermal spectrum.
Note that, for fermions, the unitarity relation is
$\f{\alpha}\cdot\f{\alpha}^\dagger+\f{\beta}\cdot\f{\beta}^\dagger=\f{1}$
instead of 
$\f{\alpha}\cdot\f{\alpha}^\dagger-\f{\beta}\cdot\f{\beta}^\dagger=\f{1}$
leading to the Fermi-Dirac spectrum $1/(e^{2\pi\omega/\kappa}+1)$ for
the usual Hawking radiation.
If the occupation number (i.e., $\ket{0}$ or $\ket{1}$) depends on the 
history of the mode -- e.g., the frequency $\omega$ -- then one would obtain 
another (in general non-thermal) spectrum.
With an appropriate mixture of $\ket{0}$ and $\ket{1}$, one could even obtain 
a state with no outgoing particle content (Hawking radiation) at all
(Boulware vacuum).

In summary, an interaction with a reservoir at the Planck scale with 
respect to the rest frame of the black hole can invalidate Hawking's 
derivation. 
In that argument, the rest frame of the black hole is a crucial point -- 
a damping term with respect to the freely falling frame 
$2i(\omega+v_{\rm fl}k)\gamma(k)$ would not induce a positive imaginary 
part of $\omega$.
A similar phenomenon occurs in the so-called `` black hole laser''
where wavepackets bounce back and forth between the inner and outer
horizons -- which also generates deviations from the Hawking effect
\cite{laser}. 
This quasi-reflection mechanism also singles out the rest frame of the 
black hole (location of the two horizons) as a preferred frame for the 
Planckian modes.
In contrast to the Miles instability, this phenomenon displays more 
similarities to the Pierce instability \cite{pierce}.

\subsection{Breakdown of Adiabaticity}\label{Breakdown}

In the previous subsection \ref{Miles}, the assumption {\bf a)} of 
Section~\ref{Universality} and hence also {\bf b)} failed. 
Let us now give an example for the breakdown of the adiabaticity
condition {\bf c)}, which is closely related to the assumption that
geometric optics is valid everywhere except in the vicinity of the
horizon. 

One version of the adiabatic theorem states that if the dynamics of all 
internal degrees of freedom is much faster than any external time
dependence, then a system being initially in its ground state
basically remains in the (time-dependent instantaneous) ground state.
(Of course, for this theorem to apply we have to assume that quantum
theory is still valid at the  Planck scale.)

As a counter-example, where the system does not stay in its ground
state, consider the dispersion relation
\bea
\omega^2=\sin^2k+m^2
\,,
\ea
which has minima at $k\in\pi\mathbb N$ (Planckian units).
If we assume a weakly time-dependent metric far away from the black
hole $ds^2=a^2(t)[dt^2-dx^2]$, the wave equation reads after a normal
mode expansion  
\bea
\ddot\phi_k+\left[\sin^2k+a^2(t)\,m^2\right]\phi_k=0
\,,
\ea
since the mass term breaks the conformal invariance.
Therefore, it very easy to create Planckian ($k\approx\pi\mathbb N$)
particles (e.g., via parametric resonance) by means of comparably
small and slow (sub-Planckian) variations of $a(t)$ with with a
characteristic scale corresponding to $m$ (instead of the Planck mass).

As a result of this breakdown of the geometric optics approximation far 
away from the horizon, the Planckian modes falling towards the black hole
are not in their ground state -- and hence one will again obtain deviations 
from the Hawking effect.
Note that a similar effect (occupation of Planckian modes) can occur
during inflation if we assume a dispersion relation like the above.

\section{Conclusions}\label{Conclusions}

\subsection{Summary}\label{Summary}

The Hawking effect is not {\em a priori} independent of the laws of physics 
at the Planck scale, but it can be made so by imposing the three assumptions 
{\bf a) Freely falling frame},
{\bf b) Ground state}, and 
{\bf c) Adiabatic evolution},
explained in more detail in Section~\ref{Universality}.
As one example, we generalized the analytical method of Ref.~\cite{analytical}
to arbitrary dispersion relations subject to some rather general assumptions.

However, we have also demonstrated counter-examples, which do not appear 
to be unphysical or artificial, displaying deviations from Hawking's result.
Therefore, whether real black holes emit Hawking radiation or not remains an 
open question and gives non-trivial information about Planckian physics.

\subsection{Outlook}\label{Outlook}

Another example, where sub-Planckian phenomena have their origin in 
trans-Planckian modes, is the generation of inhomogeneities during the
cosmic epoch of inflation (according to our present standard model of
cosmology) from quantum fluctuations of the inflaton field.
In this case, the investigation of the universality, or, conversely,
the dependence of this mechanism on Planckian physics including
higher-order corrections has an additional aspect, because
observations of the cosmic micro-wave background, for example, might
yield signatures of Planckian physics, see, e.g.,
\cite{brandenberger,inflation}. 
E.g., a dispersion relation with minima for large $k$-values as in
Section \ref{Breakdown} potentially allows particle creation leading
to a change  in the spectrum, see \cite{brandenberger}.

\acknowledgments

R.~S.~gratefully acknowledges financial support by the Emmy-Noether
Programme of the German Research Foundation (DFG) under grant 
No.~SCHU 1557/1-1 and by the Humboldt foundation 
as well as valuable conversations with G.~Volovik 
during a visit at the Low Temperature Laboratory in Finland, 
which was supported by the Programme EU-IHP ULTI 3.
Furthermore, this work was supported by the COSLAB Programme of the ESF, 
the CIAR, and the NSERC.


\end{document}